\documentclass[letter, traditabstract]{aa} 
\usepackage{graphicx}
\usepackage{txfonts}
\begin{document}
  \title{Observations of the forward scattering Hanle \\ 
effect in the Ca\, {\sc i} 4227 \AA\, line}

  \author{M.~Bianda \inst{\ref{inst1}}
        \and R.~Ramelli \inst{\ref{inst1}}
       \and L.~S.~Anusha \inst{\ref{inst3}}
        \and J.~O.~Stenflo \inst{\ref{inst1},\ref{inst2} }
       \and K.~N.~Nagendra \inst{\ref{inst3}}
       \and R.~Holzreuter \inst{\ref{inst4},\ref{inst2} }
       \and M.~Sampoorna  \inst{\ref{inst3}}
       \and H.~Frisch  \inst{\ref{inst5}}
       \and H.~N.~Smitha  \inst{\ref{inst3}}
}

  \institute{
 Istituto Ricerche Solari Locarno, Via Patocchi, 6605
    Locarno-Monti, Switzerland \label{inst1}
  \and 
 Institute of Astronomy, ETH Z\"urich, CH-8093 Z\"urich,
    Switzerland \label{inst2}
      \and 
 Indian Institute of Astrophysics, Koramangala, Bangalore 560 034,
    India \label{inst3}
      \and 
  MPI f\"ur Sonnensystemforschung, D-37191 Katlenburg-Lindau,
    Germany  \label{inst4}
      \and 
  UNS, CNRS, OCA, Laboratoire Cassiop\'ee, F-06304 Nice Cedex,
    France \label{inst5}
}

 \date{Received , 18 April 2011; accepted 5 May 2011}

\abstract
{
Chromospheric magnetic fields are notoriously difficult to measure. The
chromospheric lines are broad, while the fields are producing a
minuscule Zeeman-effect polarization. A promising diagnostic alternative is
provided by the forward-scattering Hanle effect, which can be recorded in
chromospheric lines such as the He {\sc i} 10830\,\AA\ and the Ca {\sc i}
4227\,\AA\ lines. We present a set of
spectropolarimetric observations of the full Stokes vector obtained near the
center of the solar disk in the Ca {\sc i} 4227\,\AA\ line with the ZIMPOL
polarimeter at the IRSOL observatory. We detect a number of interesting
forward-scattering Hanle effect signatures, which we
model successfully using polarized radiative transfer. Here we focus on the
observational aspects, while a separate companion paper deals with the
theoretical modeling. }



\titlerunning{Observations of the forward-scattering Hanle effect}

\authorrunning{M. Bianda et al.}

\maketitle

\section{Introduction}
Observations of the Hanle effect on the Sun are usually carried out near the
limb of the Sun, where the average scattering angles are large and as a
consequence the non-magnetic scattering polarization reaches a 
maximum. However, although the non-magnetic polarization vanishes at disk
center because of the axial symmetry of the radiation, in some spectral lines
magnetic fields can generate significant scattering polarization there through 
the so-called forward-scattering Hanle effect, as pointed out by 
 \cite{jtb01}. The first observation of this effect was obtained by \cite{jtbetal02}, 
who observed a  solar filament in the He {\sc i} 10830\,\AA\ triplet with the
Tenerife Infrared Polarimeter attached to the Vacuum Tower Telescope 
of the Observatorio del Teide (Tenerife, Spain). \cite{jtbetal02} 
were also able to determine the magnetic field vector of the observed filament 
by modeling the
Stokes $Q$ and $U$ profiles (produced by the forward-scattering Hanle effect)
and the Stokes $V$ profiles (caused by the longitudinal Zeeman effect).
Observations in filaments carried out by \cite{biandaetalspw4} in
the H {\sc i} 6563\,\AA\ and He {\sc i} 5876\,\AA\ lines showed similar
polarization signatures.

The forward-scattering Hanle effect could also be detected in the solar
chromosphere in the Ca~{\sc i}~4227\,\AA\ line by \cite{joos02} with ZIMPOL
at the IRSOL telescope. The line core is formed in the 
low  chromosphere, around a height of about 1000 km, as 
can be calculated with the quiet sun 
model-C of \cite{fontenela93} (see for example Fig. 3 
in \cite{holzretal05} calculated for Na~{\sc i} D2.  
The Ca~{\sc i} 4227\,\AA\ line gives very similar results as can
be deduced by Fig. 5, same paper). 
At these heights,
Zeeman-based diagnostics of the vector magnetic field encounters
huge difficulties because of  the weakness of the transverse Zeeman effect for
broad spectral lines. This is because the
Zeeman-effect linear polarization scales  as the square of the ratio of
Zeeman splitting to line width. Therefore we need to explore the potential of
using the forward-scattering Hanle effect 
in the Ca~{\sc i}~4227\,\AA\ line as a tool to diagnose
chromospheric vector magnetic fields near the center of the solar disk. The
 theoretical investigations of the Hanle effect on the Ca  {\sc i}
4227\,\AA\ line by  \cite{sametal09} and  \cite{anuetal10} (see also
Anusha et al.,  2011b)
were extended by \cite{anuetal11a} to include the forward-scattering
signatures (see also \cite{manso10} concerning the IR triplet of Ca\,{\sc ii}).
 Therefore a dedicated observing campaign 
 to investigate forward-scattering in
 the the  Ca~{\sc i}~4227\,\AA\ line
 was carried
out in October-November 2010 with the ZIMPOL polarimeter at IRSOL. The
theoretical analysis and interpretation of these observations is presented in
detail in \cite{anuetal11a}. 

In the present paper we describe the observational aspects of this
campaign. Section~2 explains how the observations were obtained and reduced,
Section~3 presents the observational results, and our conclusions are 
given in Secion.~4.

\section{Observations and data reduction}

Observations were carried out at IRSOL over  13 days during
October-November 2010 with the 45 cm aperture Gregory
Coud\'e telescope, the Czerny Turner spectrograph and the new ZIMPOL-3
polarimeter version (Ramelli et al., 2010). 
We collected 56 independent measurements in regions at $\mu>0.8$. Moderately 
active regions and very quiet regions were selected.
The spectrograph slit was 65 $\mu$m wide corresponding to 0\arcsec.5 on the
solar disk, and its length subtended 184\arcsec.
To select the correct order in the spectrograph, we used an
interference filter with high transmission (70\% at 4227 \AA).

\begin{figure}
\centering
\includegraphics[width=0.7\linewidth,angle=0]{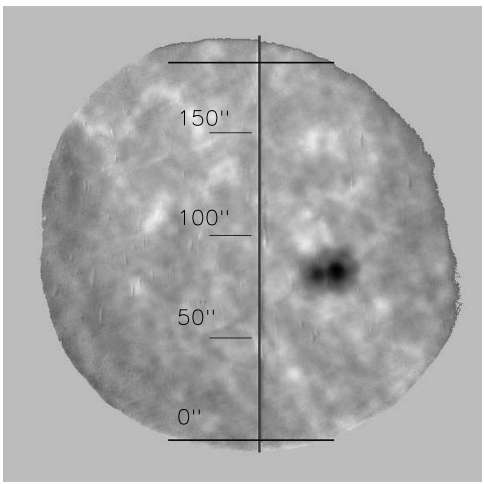}
\caption{The slit-jaw image recorded on October 21, 2010 with a $580\times 580$
  pixel CCD camera using an interference filter with 1.5 \AA\ bandwidth
  centered on the Ca {\sc ii} 3933 \AA\ K line.  The spot is in the NOAO 
  region 1113 at W20N16. The dark vertical line represents the slit, the
  dark horizontal lines delimit the 0\arcsec\ to 184\arcsec\ spatial range
  imaged on the ZIMPOL sensor, and the  spatial scale in arcsec is indicated. 
 The  slit is oriented such that it is parallel to the heliographic polar axis.}
\label{fig-1}
\end{figure}

The ZIMPOL technique is based on a fast modulation analyzer
connected to a special demodulating CCD sensor 
on which three out of four pixel rows are
masked. 
Shifting synchronously the collected photo-electron charges between the 
masked and  unmasked pixel rows, one gets four intensity images 
corresponding to different modulation phase intervals. 
The Stokes images are obtained through  
linear combinations of these four images. Owing the high modulation
frequency, seeing effects are effectively frozen and do not cause spurious 
polarization signatures in the fractional polarization images $Q/I$, $U/I$, and
$V/I$ (see Gandorfer \& Povel, 1997 for more details). 

Since we used a single piezo-elastic modulator (working at 42 kHz),  
two separate measurements were required 
to obtain the full Stokes vector. 
The two measurements are carried out by rotating
the analyzer (piezo-elastic modulator and linear polarizer) by 
$45^{\circ}$. In the first position, we obtain the Stokes $I$, $Q/I$, and 
$V/I$ determinations, and in
the second position $I$, $U/I$, and $V/I$.
A sequence of 100 frames each exposed for 1 second are recorded at each
position.  
In total, we recorded 5 or 10 sequences of 100 frames per position.
The total time required for 10 sequences is about one hour. 

With the help of an image rotator (a Dove prism), the solar image
on the spectrograph slit plane is rotated to the desired orientation.
The image rotation caused by the Gregory-Coud\'e
telescope is compensated automatically. 
The polarization analyzer is
positioned in front of the Dove prism and rotates accordingly.

The image on the spectrograph slit plane (slit jaw image) is re-imaged by a
telecentric system on a CCD camera (Figure~\ref{fig-1}). 
A 1.5 \AA\ FWHM Daystar interference filter 
centered on the Ca {\sc ii}  K  3933 \AA\  line is placed in front of the
slit-jaw camera  that allows us to locate the measured region on the 
Sun and to more clearly identify relevant structures in the selected
region. Slit-jaw images are recorded every eight seconds to verify the
accuracy of the tracking.

A typical observing run includes dark and flat field measurements as well as
polarization calibrations, which are obtained with  
linear and circular polarizers that are inserted near the exit window of the
telescope. 

We cannot insert the calibration optics before this location. A glass
tilt-plate, whose 
orientation is controlled by stepper motors, is inserted just
after the calibration optics and allows us to compensate for the linear
polarization that is caused by the telescope.
The remaining instrumental polarization effects produced by the telescope are
corrected for in the data reduction.

The position of the observed region is kept fixed on the spectrograph slit by
the automatic guiding system 
(K\"uveler et. al, 1998, 2010, 2011)
which 
{also compensates} for the solar rotation. The position is held with a precision
of 1\arcsec.5 to 2\arcsec\ (RMS).

\begin{figure}
\centering
\includegraphics[width=\linewidth,angle=0]{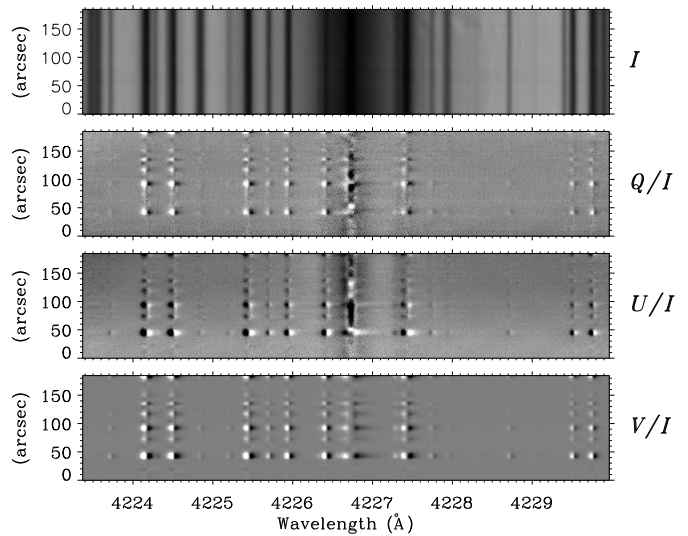}
\caption{Stokes images in a spectral window around the Ca {\sc i} 4227 \AA\,
  line. The observation was performed in the region shown in
  Fig.~\ref{fig-1}. Dark current and polarization calibration
  were taken into account, but not the instrumental cross talks between the
  Stokes parameters. 
In the $Q/I$ and $U/I$ images, the grey scale cuts are $\pm 0.2\%$, while in the 
$V/I$ image they are $\pm 2.5\%$.}
\label{fig-2}
\end{figure}

Figure~\ref{fig-2} shows polarization images obtained on 
October 21, 2010 near the sunspot in NOAO region 1113.  Dark recordings and
polarization calibrations were taken into account, and spikes caused by
cosmic rays or hot pixels have been removed by a software filter.
Flat field and Fourier filter corrections were applied additionally.
The corresponding slit position is shown
in Fig.~\ref{fig-1}, with the slit oriented parallel to the
heliographic polar axis. The center of the spectrograph slit is
located at N17 W18.
The spectral range
covers 6.6~\AA\ distributed over 1240 pixels (5.3 m\AA\ per pixel). In the
spatial direction an angular extent of 184\arcsec\ is covered by 140
(unmasked) pixel rows (thus  
effectively 1\arcsec.3 per pixel).
The typical statistical pixel noise in the polarization images is about 
$2\times 10^{-4}$. 

The observations were carried out in October and November 2010 when
the instrumental 
effects were no longer  negligible, 
the main relevant source of troubles beeing the 
circular to linear polarization cross talk.
The correction
 coefficients are extracted from the instrumental polarization measurements
shown in Figures 1 and 2 of  \cite{ramellietal05}.

\begin{figure}
\centering
\includegraphics[width=8cm,angle=0]{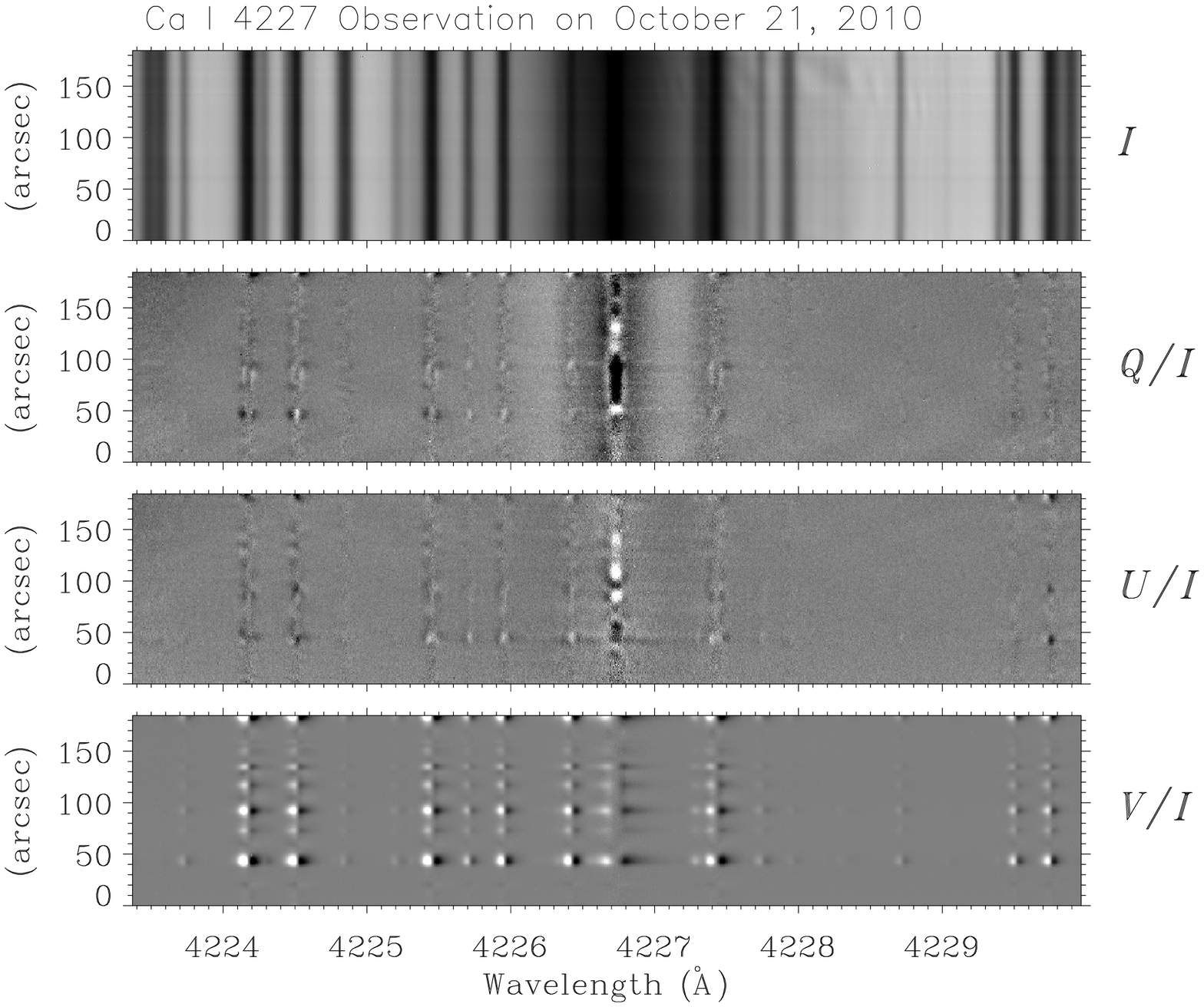}
\includegraphics[width=8cm,angle=0]{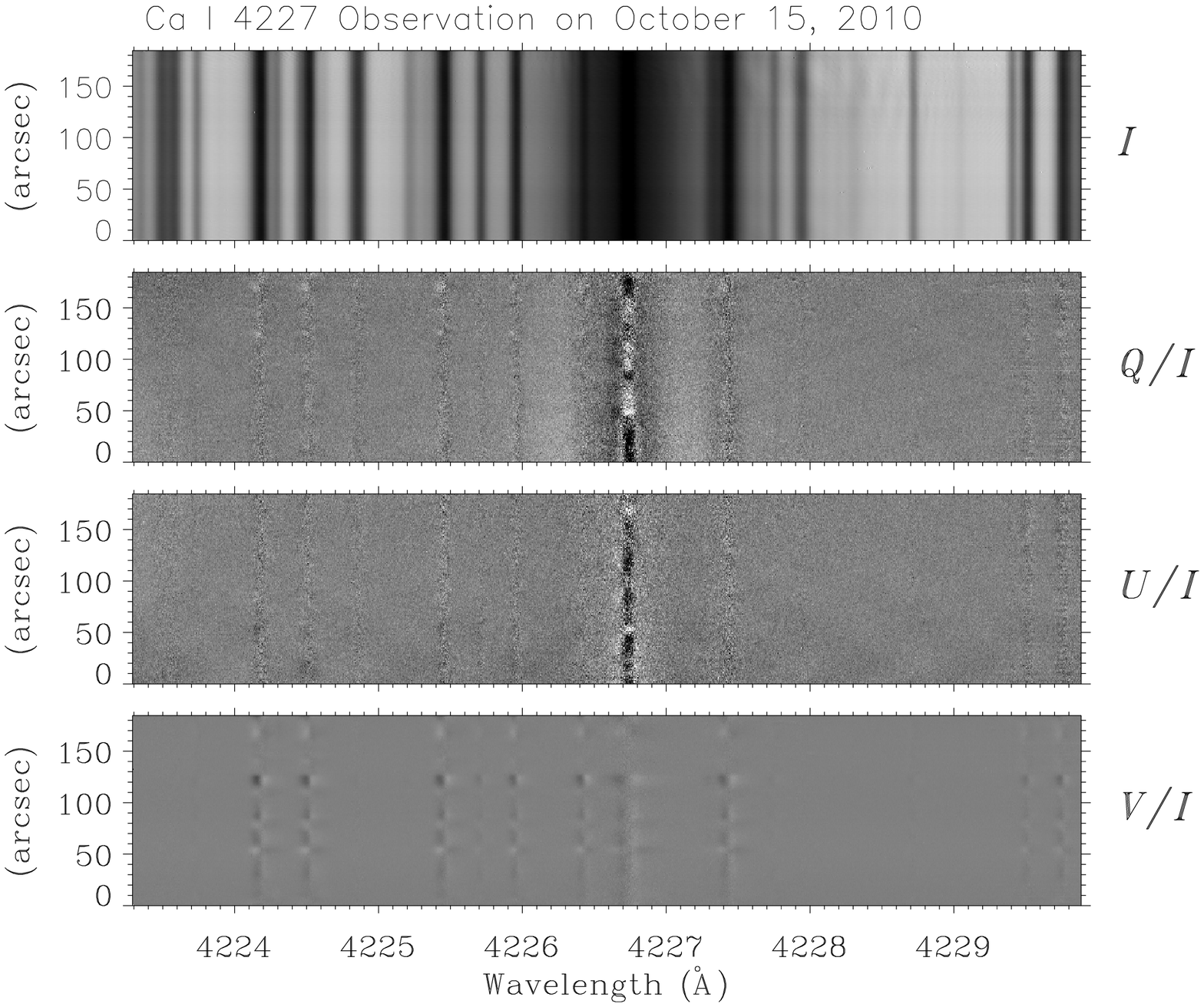}
\includegraphics[width=8cm,angle=0]{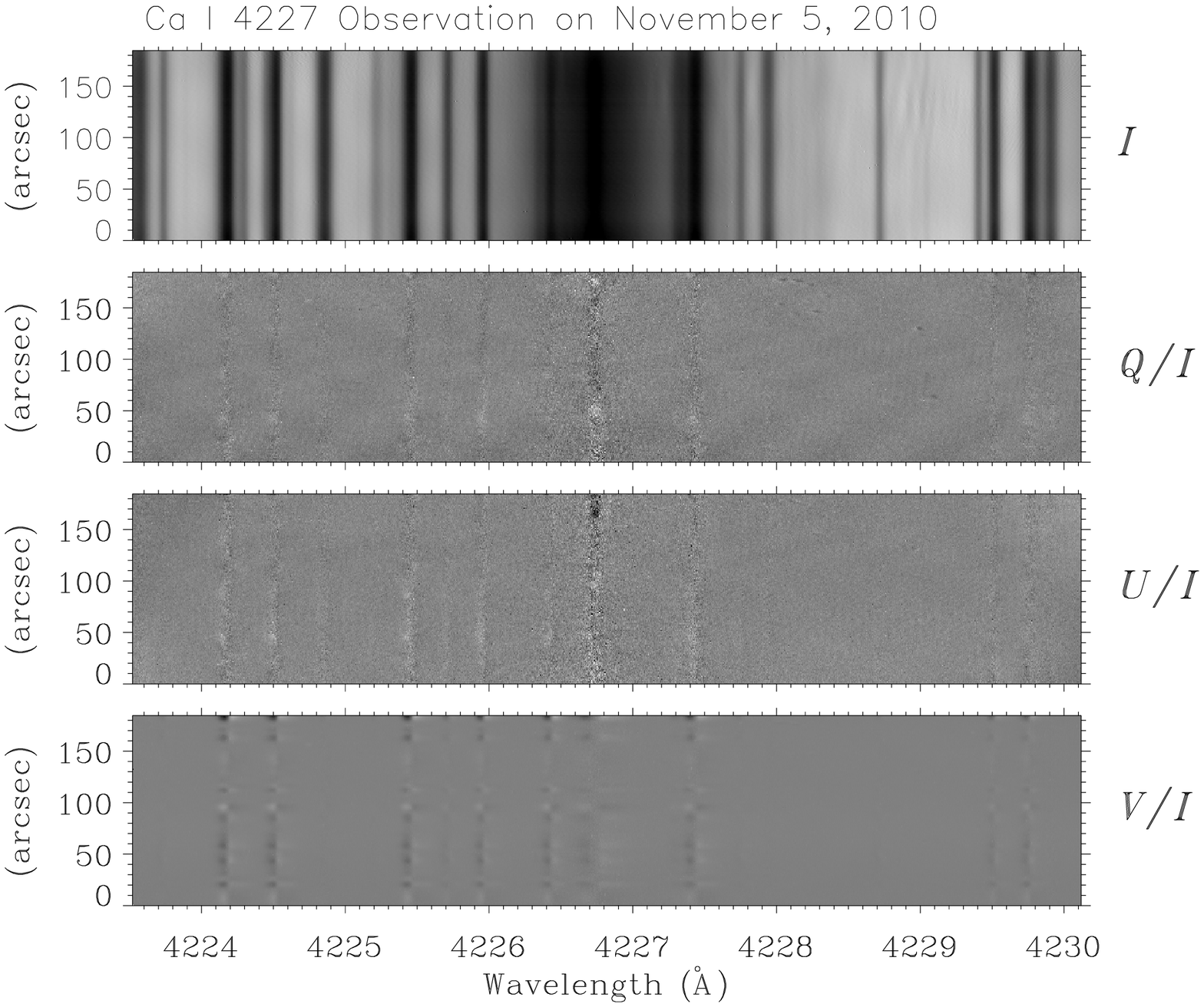}
\caption{Top observation: the same recording as in  Figure~\ref{fig-2}
    after application of the cross-talk correction and of the linear
    polarization basis rotation (see text). The measurement was done around $\mu=0.94$.
The second observation (middle) reports measurements of
October 15 (around $ \mu=0.95$) in the region shown in  Fig.~\ref{fig-5}.
The third observation was done on November 5 (bottom) very
close to disk center ( $ \mu=1$) in a very quiet region. Here we note the absence of linear
polarization signature.  
For all recordings, the grey scale cuts are the same as in Fig.~\ref{fig-2}.}
\label{fig-3}
\end{figure}

Positive $Q/I$ is defined as the amount of 
linear polarization along the spectrograph
slit. To convert the Stokes images to the linear 
polarization basis
used for the theoretical model (Anusha et al., 2011a),
a basis transformation is
needed.  
In the new basis,  the positive $Q/I$ direction is defined to be perpendicular to the local
radius vector for each point along the
spectrograph slit. 
The basis transformation thus has to be applied pixel by pixel along the slit,
using the angle formed by the slit with the local radius vector. 
The top panel in Fig.~\ref{fig-3} shows the results after these corrections. 

\section{Results}

Among the observations obtained during this observing campaign,
the results illustrated in the  top and middle panels of
Fig.~\ref{fig-3}
are typical of observations performed close
to moderately active regions. 
We first consider the $Q/I$
and $U/I$ signatures in the Ca {\sc i} 4227 \AA\, line center. 

As shown by \cite{anuetal11a}, the observed amplitude of the linear
polarization at line center is much too large to be explainable in terms of
the transverse Zeeman effect, but can be interpreted and modeled in terms
of the forward-scattering Hanle effect. 

The observation of October 21  was performed 
rather close to disk center. 
The $\mu$ ($=\cos\theta$, where $\theta$ is the heliocentric angle)
value varies along the slit from 
$\mu=0.96$ at coordinate 0\arcsec\ to $\mu=0.915$ at coordinate 184\arcsec.
Even at these quite large  $\mu$ values, we can recognize weak features of
about $3\times 10^{-4}$ in
$Q/I$ produced by scattering-polarization in the line wings 
around 4226.2~\AA\ and 4227.1~\AA,  arising from
the so called partial frequency redistribution mechanism in resonance scattering 
(see Anusha et al., 2011b).

\begin{figure}
\centering
\includegraphics[width=\linewidth]{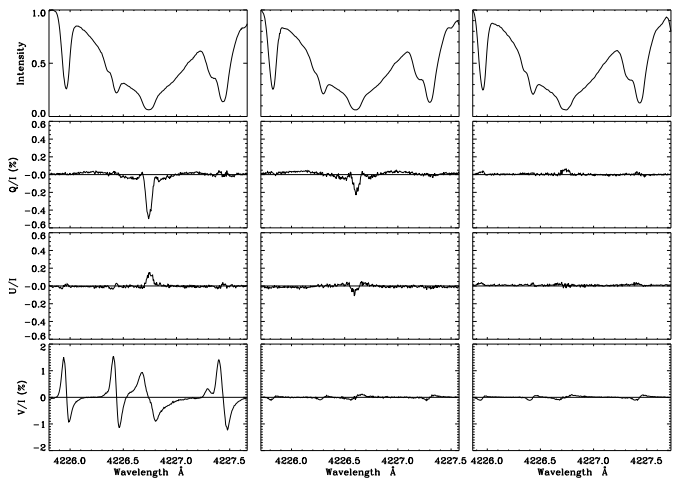}
\caption{Stokes spectra extracted from the observations in
  Fig.~\ref{fig-3}. Left panels: data of October 21 at coordinate 88\arcsec.   
Central panels data measured on October 15, around coordinate 25\arcsec .
Right panels: Stokes profiles recorded in the very quiet 
region near disk center on November 5 around coordinate  52\arcsec.}
\label{fig-4}
\end{figure}

The blend lines, most of them Fe {\sc i} lines, show the usual Zeeman
signatures caused by oriented magnetic fields in the layers where 
they are formed (see $V/I$ profiles in the lower
panels of Fig.~\ref{fig-4}).

In Figure~\ref{fig-4}, the  Stokes profiles extracted from
the Stokes images reported in
Fig.~\ref{fig-3} are presented. In the left panel,
the profiles were averaged over four pixels, corresponding to 5\arcsec.3
along the spatial direction, at slit coordinate 88\arcsec. The amplitudes of the
linear polarization peaks in the left panels are among the largest that we
found in our observing campaign. The largest  degree of linear polarization
that was recorded was about 0.75\,\% (not shown here).

The central panels show Stokes profiles extracted from data measured on 
October 15,  averaged over 12\arcsec around coordinate 25\arcsec.  
Data measured in this region were analyzed by \cite{anuetal11a}.
The slit was placed near the active region NOAO 1112, as shown in the
magnetogram 
in Fig. 5, for which the grey scale cuts are at $\pm100$ G. 
We can  see that although the local photospheric magnetic field is faint at the
slit position, the forward-scattering Hanle effect gives clear linear 
polarization signatures there.

The Stokes profiles in the right panels refer to the recording on 
November 5 
done in a very quiet region near disk center,  
averaged over 9\arcsec\ around coordinate 52\arcsec.
The chosen portion of the slit
represents the place with a detectable linear 
polarization signature.  In really quiet regions far from
(moderately) active regions, it is relatively rare to find polarization
signatures that are significantly above the noise level.

To measure a significant forward-scattering signature, we need a
magnetic field with a well defined azimuth direction within the resolution
area observed (see for instance Fig. 1 in Trujillo Bueno, 2001). 
The absence of signatures can be interpreted in terms of either 
turbulent magnetic structures, or very faint magnetic structures, or 
fast evolving structures in the observed area.
As the total observation time per recording is of the order one hour, we
average over effects such as temporal evolution and the local proper motions
of the atmospheric structures. Nevertheless, the well-defined  quality of the
polarization signatures, and both their profile shapes and spatial distributions
indicate that long-lived chromospheric magnetic structures with spatial scales
of order 5\arcsec\ exist, and that we can therefore use this kind of data to
diagnose chromospheric vector magnetic fields in the neighborhood of
moderately active regions.

\begin{figure}
\centering
\includegraphics[width=6cm]{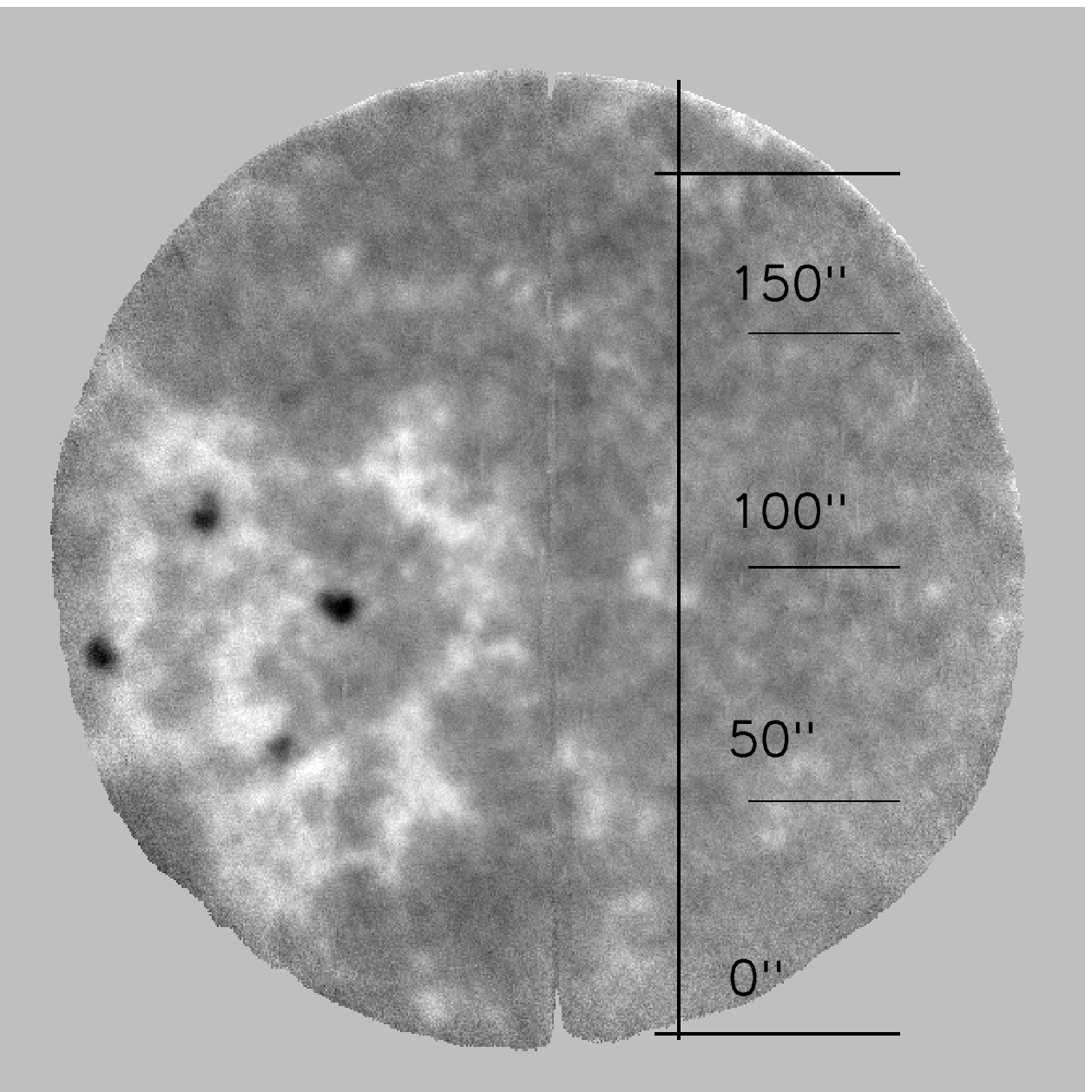}
\includegraphics[width=6cm]{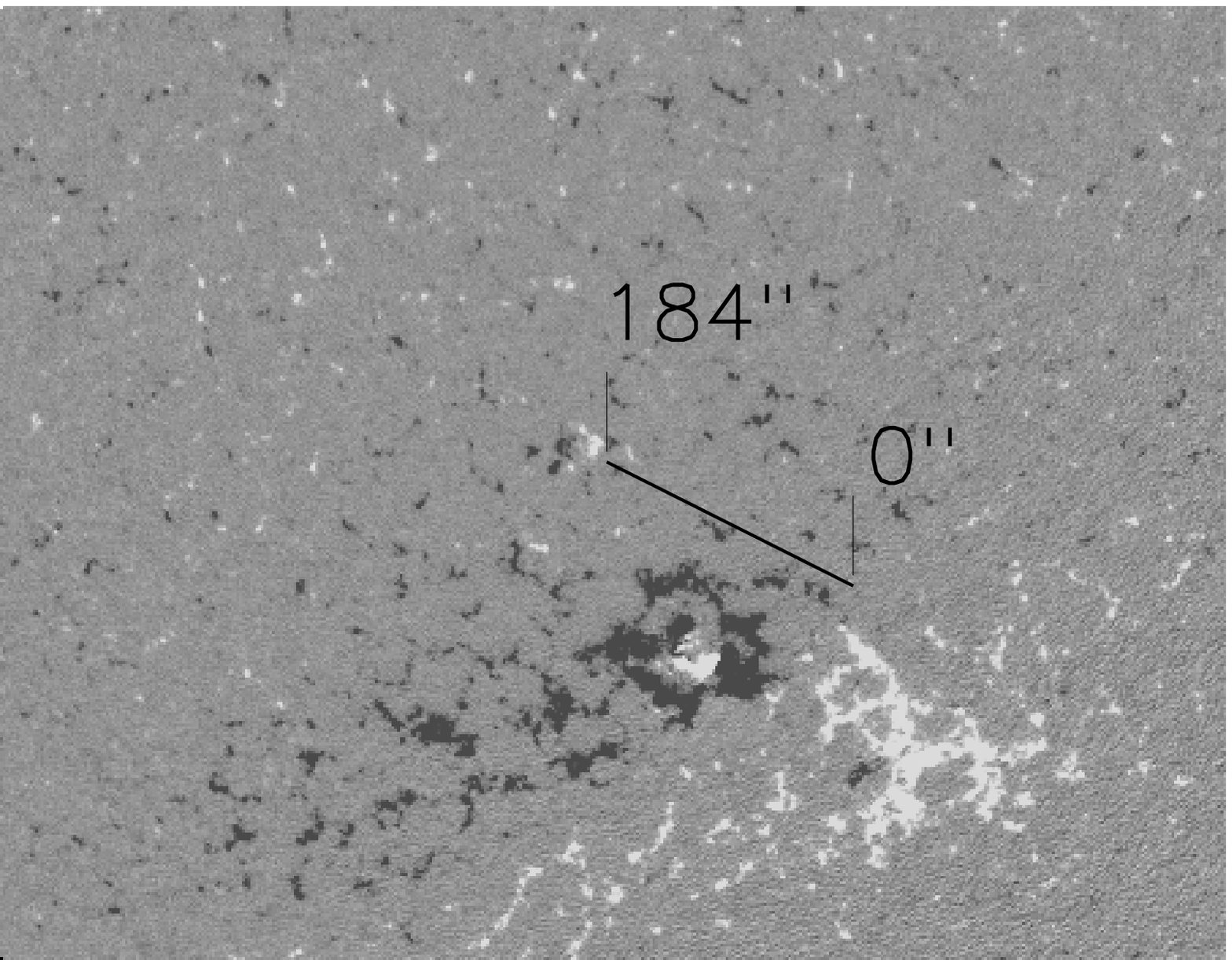}
\caption{Top panel: the slit jaw image of the region observed on October 15,
  2010.  Bottom panel: a  $900\arcsec \times 700\arcsec$  area of an MDI
  magnetogram recorded on October 15, 2010, approximately at the same time as
  our 
  observation. Heliographic north is up. Grey scale cuts at $\pm100$ G
  are used. Overplotted is the position of the spectrograph slit.
}
\label{fig-5}
\end{figure}

\section{Conclusions}

Our observations of the forward-scattering Hanle effects in the Ca {\sc i}
4227\,\AA\ line has revealed many well-defined polarization signatures that could
be reproduced with synthetic Stokes profiles computed with coherent scattering
in  one-dimensional semi-empirical models of the solar 
atmosphere (Anusha et al., 2011a).
We recorded polarization profiles with amplitudes up to about 0.7\,\% in
the degree of linear polarization, which is much higher than can be explained
in terms of the transverse Zeeman effect. The signal-to-noise ratio level of about   
$2\times 10^{-4}$ per pixel is reached in about one hour
 with a 45 cm aperture telescope.
The richness of the profile shapes and the significant spatial variations on a
scale of a  
few arcsec show that long-lived chromospheric magnetic structures are common.
This together with the accompanying theoretical analysis of \cite{anuetal11a}
demonstrates that the forward-scattering Hanle effect in the Ca {\sc i}
4227\,\AA\ line is indeed a useful and important 
tool to measure the vector magnetic field (amplitude and orientation) in the
low chromosphere near the center of the solar disk.

\begin{acknowledgements}
IRSOL is financed by Canton Ticino and the city of Locarno, 
together with the municipalities affiliated with CISL.
The project has been supported by the Swiss Nationalfonds (SNF) grant
200020-127329. 
L.S.A. thanks the Indo-Swiss Joint Research Program
(ISJRP) and IRSOL for supporting her visit to IRSOL. 
R.R. acknowledges the financial support provided by the foundation Carlo e
Albina Cavargna.  
IRSOL acknowledges the financial support provided by the foundation Aldo e
Cele Dacc\`o.
The ZIMPOL electronics is now developed at SUPSI under the direction of
I. Defilippis. 
We are grateful to D. Gisler for his decisive help in the ZIMPOL project.
The improved version of the Primary Image Guiding system has been developed at
RheinMain University of Applied Sciences under the direction of G. K\"uveler. 
We are greateful to an anomymous referee for the useful suggestions.
\end{acknowledgements}

\eject

\end{document}